\documentclass[twocolumn,showpacs,amsmath,amssymb,prb,superscriptaddress]{revtex4}

\usepackage{graphicx}
\usepackage{bm}
\begin{document}

\title{Surface superconductivity in multilayered rhombohedral graphene: Supercurrent}

\author{ N.B.\ Kopnin } \affiliation{ Low
Temperature Laboratory, Aalto University, P.O. Box 15100, 00076
Aalto, Finland} \affiliation{ L.~D.~Landau Institute for
Theoretical Physics, 117940 Moscow, Russia}

\date{\today}

\begin{abstract}
The supercurrent for the surface superconductivity of a flat-band
multilayered rhombohedral graphene is calculated. Despite the
absence of dispersion of the excitation spectrum, the supercurrent
is finite. The critical current is proportional to the
zero-temperature superconducting gap, i.e., to the superconducting
critical temperature and to the size of the flat band in the
momentum space.
\end{abstract}

\maketitle

\section{Introduction}

Fermionic systems with dispersionless branches of excitation
spectrum (flat bands) have quite unusual properties; nowadays they
attract lots of research interest. Flat bands were predicted in
many condensed matter systems, see for example
Refs.~\cite{Khodel1990,NewClass,Shaginyan2010,Gulacsi2010}. In
some cases the flat bands are protected by topology in momentum
space; they emerge  on the surfaces of gapless topological
matter\cite{HeikkilaKopninVolovik10} such as surfaces of nodal
superconductors \cite{Ryu2002,SchnyderRyu2010}, graphene edges
\cite{Ryu2002}, surfaces of multilayered graphene structures
\cite{GuineaCNPeres06,MakShanHeinz2010,HeikkilaVolovik10-1}, and
in the cores of quantized vortices in topological superfluids and
superconductors
\cite{KopninSalomaa1991,Volovik2011,HeikkilaKopninVolovik10}.

The singular density of states (DOS) associated with the
dispersionless spectrum may essentially enhance the transition
temperature opening a new route to room-temperature
superconductivity. The corresponding critical temperature depends
linearly on the pairing interaction strength and can be thus
considerably higher than the usual exponentially small critical
temperature in the bulk
\cite{Khodel1990,HeikkilaKopninVolovik10,KopninHeikkilaVolovik11}.
It was shown in
\cite{HeikkilaKopninVolovik10,KopninHeikkilaVolovik11} that the
flat band that appears on the surface of multilayered rhombohedral
graphene is especially favorable for surface superconductivity.
Formation of surface superconductivity is enhanced already for a
system having $N\geq 3$ layers, where the normal-state spectrum
has a slow power-law dispersion $\xi_{p} \propto |{\bf p}|^N$ as a
function of the in-plane momentum ${\bf p}$. The DOS $\nu(\xi_{
p}) \propto \xi_{p}^{(2-N)/N}$ has a singularity at zero energy
which results in a drastic enhancement of the critical
temperature.

Absence of dispersion in a flat band raises the questions of
superconducting velocity and of the supercurrent: Can they be
nonzero and, if they can, what is then the magnitude of the
critical current? In this Letter we address the problem of
supercurrent associated with the surface superconductivity in the
flat-band multilayered rhombohedral graphene. Based on the model
employed in Ref. \cite{KopninHeikkilaVolovik11} for description of
the surface superconductivity we calculate the supercurrent as a
response to a small gradient of the order parameter phase using an
approach similar to that used for calculations of the supercurrent
in a single layer of graphene \cite{KopninSonin10}. We demonstrate
that the supercurrent is finite; the critical current is
proportional to the superconducting zero-temperature gap, i.e., to
the critical temperature, and to the radius of the flat band in
the momentum space. Being produced by the surface
superconductivity, the total current through the sample is
independent of the sample thickness.

\section{The model}

As in Ref. \cite{KopninHeikkilaVolovik11} we consider multilayered
graphene structure of $N$ layers in the discrete representation
with respect to interlayer coupling. For simplicity we choose the
rhombohedral stacking configuration considered in
\cite{GuineaCNPeres06,MakShanHeinz2010,HeikkilaVolovik10-1,HeikkilaKopninVolovik10,KopninHeikkilaVolovik11}
and assume that the most important are hoppings between the atoms
belonging to different sublattices parameterized by a single
hopping energy $t$. More general form of the multilayered
Hamiltonian can be found in
Refs.~\cite{McClure69,CastroNeto-review}. In the superconducting
case the Hamiltonian has the form of a matrix in the Nambu space.
The Bogoliubov--de Gennes (BdG) equations are
\[
\sum_{j=1}^N \left( \begin{array}{cc} \hat H_{ij}-\mu \delta_{ij}
&
\Delta_i \delta_{ij} \\
\Delta_i^*\delta _{ij} & -\hat H_{ij}+\mu \delta_{ij}
\end{array}\right)\left(
\begin{array}{c} \hat u_j \\ \hat v_j\end{array}\right)=E \left(
\begin{array}{c} \hat u_i \\ \hat v_i\end{array}\right),
\]
where the sum runs over the layers. The normal-state Hamiltonian
\cite{HeikkilaVolovik10-1}
\begin{equation}
\hat H_{ij}=v_F(\hat {\bm \sigma}\cdot {\bf p})\delta_{i,j}-t\hat
\sigma_+ \delta_{i,j+1} -t\hat \sigma_- \delta_{i,j-1}\ ,
\label{H-norm}
\end{equation}
$\hat{\bm \sigma}=(\hat\sigma_x,\ \hat\sigma_y)$, $\hat \sigma_\pm
= (\hat \sigma_x \pm i\hat \sigma_y)/2$, and $\hat u_i,\ \hat v_i$
are matrices and spinors in the pseudo-spin space associated with
two sublattices. This Hamiltonian acts on the envelope function of
the in-plane momentum ${\bf p}$ taken near one of the Dirac
points, i.e., for $|{\bf p}|\ll \hbar /a$ where $a$ is the
interatomic distance within a layer; $v_F =3 t_0 a/2\hbar $ where
$t_0$ is the the hopping energy between nearest-neighbor atoms
belonging to different sublattices on a layer. The particle-like,
$\hat u_i$, and hole-like, $\hat v_i$, wave functions near the
Dirac point are coupled via the superconducting order parameter
$\Delta_i$ that can appear in the presence of a pairing
interaction. Here we do not specify the nature of the pairing. It
can be due to either electron-phonon interaction or other pairing
interactions that have been suggested as a source for intrinsic
superconductivity in graphene, see Refs.~\cite{pairing}. The
excitation energy for particles and holes is measured upwards or
downwards, respectively, from the Fermi level which can be shifted
with respect to the Dirac point due to doping. Here we assume that
the shift is the same on all layers. The order parameter and the
Fermi level shift $\mu$ are scalars in the pseudo-spin space. We
assume that $\Delta$ and $\mu$ are much smaller than the
inter-layer coupling energy $t>0$, which in turn is $t\ll t_0$.
Usually, $t\sim 0.1\, t_0$ where $t_0\sim 3$ eV
\cite{CastroNeto-review}.

We decompose the wave function
\begin{equation}
\left(\begin{array}{c} \hat u_n\\ \hat v_n
\end{array}\right)=\left[ \left(\begin{array}{c} \alpha_n^+
\\ \beta_n^+ \end{array}\right) \otimes \hat \Psi^+  +
\left(\begin{array}{c} \alpha_n^-
\\ \beta_n^- \end{array}\right) \otimes \hat \Psi^- \right]\label{wavefunc-u}
\end{equation}
into the spinor functions localized at each sublattice
\[
\hat \Psi^+   =\left(\begin{array}{c} 1\\0\end{array}\right)
 \ , \; \hat \Psi^-   =\left(\begin{array}{c}
0\\1\end{array}\right)  \ .
\]
We introduce matrices and vectors in the Nambu space
\[
\check \tau_3=\left(\begin{array}{lr} 1 & 0 \\ 0&
-1\end{array}\right)\ , \; \check \Delta
_n=\left(\begin{array}{lr} 0 & \Delta_n \\ \Delta^*_n &
0\end{array}\right)\ , \; \check \alpha_n^\pm =
\left(\begin{array}{c} \alpha_n^\pm
\\ \beta_n^\pm  \end{array}\right)\ .
\] The BdG equations take the form
\begin{eqnarray}
\check \tau_3\left[ v_F(\hat p_x-i\hat p_y)\check \alpha_n^-
-t\check \alpha_{n-1}^- -\mu \check \alpha_n^+\right]=E \check
\alpha_n^+
 ,\; n\ne 1, \quad \label{BdGeqD}\\
\check \tau_3 \left[ v_F(\hat p_x+i\hat p_y)\check \alpha_n^+
-t\check \alpha_{n+1}^+ -\mu \alpha_n^- \right]=E \check
\alpha_n^-  ,\; n\ne N, \quad \label{BdGeqA}
\end{eqnarray}
where $\hat {\bf p}$ is the momentum operator. In
Eqs.~(\ref{BdGeqD}) and (\ref{BdGeqA}) we assume that $\Delta_n
\ne 0$ only at the outermost layers, while $\Delta_n =0$ for $n\ne
1,N$. The arguments supporting this assumption are given in
Ref.~\cite{KopninHeikkilaVolovik11}; it was shown that the order
parameter quickly decays as a function of the distance from the
surface. We also neglect $\Delta_n$ as compared to $t$ in
Eqs.~(\ref{BdGeqD}) and (\ref{BdGeqA}) for $n=N$ and $n=1$,
respectively, as they lead to higher-order corrections in
$\Delta/t$. The particle and hole channels are thus decoupled if
$n\ne 1,N$. Expanding the coefficients in plane waves $\alpha,
\beta \propto e^{i{\bf p}{\bf r}+ip_z z}$ we find the energy in
terms of in-plane ${\bf p}$ and transverse momentum $p_z$ ($d$ is
the interlayer distance) \cite{HeikkilaVolovik10-1}
\begin{equation}
E^2=v_F^2 p^2 -2tv_Fp \cos(p_z d -\phi)+t^2 \label{E-bulk}
\end{equation}
where $p=\sqrt{p_x^2+p_y^2}$ and $e^{i\phi}=(p_x+ip_y)/p$.
Equations (\ref{BdGeqD}) and (\ref{BdGeqA}) determine the
coefficients \cite{HeikkilaVolovik10-1,KopninHeikkilaVolovik11}
\begin{eqnarray}
\check \alpha_n^+=\zeta^+_n({\bf p}) \check A^+ +\zeta^-_n({\bf
p}) t^{-2}(\check \tau_3 \tilde E +\tilde \mu)v_F(p_x-ip_y)\check
A^- , \label{alpha+n} \\
\check \alpha_n^-=\zeta^-_n({\bf p}) \check A^- +\zeta^+_n({\bf
p}) t^{-2}(\check \tau_3 \tilde E +\tilde \mu)v_F(p_x+ip_y)\check
A^+ , \label{alpha-n}
\end{eqnarray}
where the basis functions are
\begin{eqnarray*}
\zeta^+_n({\bf p})=\left[v_F(p_x+ip_y)/t\right]^{n-1}\ , \\
\zeta^-_n({\bf p})=\left[v_F(p_x-ip_y)/t\right]^{N-n}\ .
\end{eqnarray*}
Here we include the first-order corrections in energy. Having an
imaginary momentum $p_z$ for $v_Fp<t$, these solutions decay away
from the surfaces  and thus they describe the surface states.
Normalization requires
\[
d\sum_{n=1}^N \left[ (\check \alpha^{+}_n)^\dagger \check
\alpha^+_n + (\check \alpha^{-}_n)^\dagger  \check
\alpha^-_n\right] =1\ .
\]
This gives
\begin{equation}
d\left[( \check A^{+})^\dagger \check A^+ +(\check A^{-})^\dagger
\check A^-\right] = 1-v_F^2p^2/t^2 \ . \label{normalization1}
\end{equation}

A finite order parameter $\Delta$ couples the particle and hole
channels at the outermost layers, $i=1$ and $i=N$,
\begin{eqnarray}
\check \tau_3 v_F(\hat p_x-i\hat p_y)\check \alpha_1^- -\check
\tau_3 \mu_1\check \alpha_1^+ &=&E \check \alpha_1^+
-\check \Delta \check \alpha_1^+ \label{eqn-u=1}\ , \\
\check \tau_3 v_F(\hat p_x+i\hat p_y)\check \alpha_N^+ -\check
\tau_3 \mu_N\check \alpha_N^- &=&E \check \alpha_N^- -\check
\Delta \check \alpha_N^- \ . \label{eqn-u=N}
\end{eqnarray}

The boundary conditions (\ref{eqn-u=1}), (\ref{eqn-u=N}) select
$p_z$ and determine $2N$ particle and hole branches of the energy
spectrum. Looking for the branches that belong to the surface
states with energies of the order of $\Delta$ and $\mu$, we solve
these equations for $E\ll t$.

\section{Supercurrent}

The operator of current along a layer couples the states at
different sublattices, $ \hat u_{ \gamma,{\bf p}}^\dagger(n) \hat
{\bm \sigma}\hat u_{ \gamma,{\bf p}} (n)+ \hat v_{q,{\bf
p}}^\dagger(n) \hat {\bm \sigma}\hat v_{\gamma,{\bf p}}(n) $. For
example, the $x$ component of current at layer $n$ is
\begin{eqnarray}
j_x(n)=-ev_F \sum_{\gamma ,{\bf p}} \left[ \check
\alpha_{\gamma,n}^{+\dagger}({\bf p}) \check
\alpha_{\gamma,n}^-({\bf p}) + \check
\alpha_{\gamma,n}^{-\dagger}({\bf p}) \check
\alpha^+_{\gamma,n}({\bf p})\right]\nonumber\\
\times \left[1 - 2f_{\gamma,{\bf p}}\right].\quad
\label{currentN}
\end{eqnarray}
where $\gamma$ labels different states for given ${\bf p}$,  while
$f_{\gamma,{\bf p}}$ is the distribution function.

To calculate the supercurrent we use the same approach as in Ref.
\cite{KopninSonin10}. Consider $\Delta=|\Delta|e^{i{\bf k}{\bf
r}}$. Separating the order-parameter phase, we put $u_n= u_n({\bf
p}) e^{i({\bf p}+{\bf k}/2){\bf r}}$, while $v_n = v_n({\bf p})
e^{i({\bf p}-{\bf k}/2){\bf r}}$. For large $N\gg 1$ the most
important corrections come from $({\bf p}\pm {\bf k}/2)^N$. (The
exact condition for $N$ will be established later.) We have
\begin{eqnarray}
\check \alpha_n^+=\zeta^+_n(\check {\bf p}) \check A^+
+\zeta^-_n(\check {\bf p}) t^{-2}(\check \tau_3 \tilde E +\tilde
\mu)v_F(p_x-ip_y)\check
A^-  , \quad \label{alpha+n-k} \\
\check \alpha_n^-=\zeta^+_n(\check {\bf p}) \check A^-
+\zeta^+_n(\check {\bf p}) t^{-2}(\check \tau_3 \tilde E +\tilde
\mu)v_F(p_x+ip_y)\check A^+  , \quad \label{alpha-n-k}
\end{eqnarray}
where $\check {\bf p}= {\bf p}+\check \tau_3{\bf k}/2$. Equations
(\ref{eqn-u=1}), (\ref{eqn-u=N}) at the outermost layers give
\begin{eqnarray}
\check \tau_3 \xi^-_{{\bf p}+\check \tau_3{\bf k}/2} \check A^-=
(\tilde E+\check \tau_3 \tilde \mu )\check A^+ - \check \tau_1
|\Delta|  \check A^+ \ ,\label{BdGequ1}\\
\check \tau_3 \xi^+_{{\bf p}+\check \tau_3{\bf k}/2} \check A^+=
(\tilde E+\check \tau_3 \tilde \mu )\check A^- - \check \tau_1
|\Delta | \check A^- \ . \label{BdGequ2}
\end{eqnarray}
Here
\[
\xi_{\bf p}^\mp =t\left[v_F(p_x\mp ip_y)/t\right]^N= e^{\mp
iN\phi}\xi_{ p}\ , \; \xi_p=t(v_Fp/t)^N
\]

Using the spinors in the sublattice space, Eqs. (\ref{BdGequ1}),
(\ref{BdGequ2}) can be written as
\begin{equation}
\left[ \check {\hat H}_0+\check {\hat H}_1\right]\check{\hat
\psi}=\tilde E \check {\hat \psi} \ , \;
\check{\hat \psi} = \left(\begin{array}{c}\check A^+\\
\check A^-\end{array}\right)\ , \label{matrix-eq}
\end{equation}
where
\begin{eqnarray}
\check {\hat H}_0&=&\check \tau_3 e^{-i\hat \sigma_z N \phi}\hat
\sigma_x \xi_{p} -\check \tau_3 \tilde \mu +\check \tau_1
|\Delta|\ , \label{H0}\\
\check {\hat H}_1&=&e^{-i\hat \sigma_z(N-1) \phi} \frac{(\hat {\bm
\sigma}{\bf k})}{2} \frac{d\xi_{p}}{d p}\ . \label{H1}
\end{eqnarray}

In the zero order in ${\bf k}$ the coefficients $ \check{\hat
\psi}^{(0)} $ satisfy
\begin{eqnarray}
\check {\hat H}_0 \check{\hat \psi}^{(0)}=\tilde E^{(0)} \check
{\hat \psi}^{(0)} \ . \label{BdGequ0}
\end{eqnarray}
The equation has four solutions
\begin{eqnarray}
\check{\hat \psi}_{1,2}=\left(\begin{array}{c} \check A_{1,2} e^{-iN\phi /2} \\
\check A_{1,2} e^{+iN\phi /2}\end{array}\right) \ , \; \tilde E_{1,2}=\pm \tilde E_0^+ \ ,
\\
\check{\hat \psi}_{3,4}=\left(\begin{array}{c} \check A_{3,4} e^{-iN\phi /2} \\
-\check A_{3,4} e^{+iN\phi /2}\end{array}\right) \ , \; \tilde
E_{3,4}=\pm \tilde E_0^- \ .
\end{eqnarray}
Here
\[
\tilde E_0^+=\sqrt{\left( \xi_{ p} - \tilde \mu \right)^2
+|\Delta|^2}\ , \; \tilde E_0^-=\sqrt{\left(\xi_{ p}+ \tilde \mu
\right)^2 +|\Delta|^2} \ ,
\]
and
\begin{eqnarray*}
\check A_1=\frac{C}{\sqrt{2}}\left(\begin{array}{c} u_+
\\ v_+ \end{array}\right)\ ,
\; \check A_2=\frac{C}{\sqrt{2}}\left(\begin{array}{c}v_+ \\
- u_+\end{array}\right)\ , \\
\check A_3=\frac{C}{\sqrt{2}}\left(\begin{array}{c} v_-  \\ u_-
\end{array}\right)\ , \; \check A_4=\frac{C}{\sqrt{2}}\left(\begin{array}{c} u_- \\
-v_-
\end{array}\right)\ .
\end{eqnarray*}
Normalization is determined by Eq. (\ref{normalization1}),
$|C|^2=d^{-1}(1-v_F^2p^2/t^2)$, the coherence factors are
\begin{eqnarray*}
u_\pm=\frac{1}{\sqrt{2}}\left[1+\frac{\xi_p \mp\mu}{\tilde
E_0^\pm}\right]^{1/2} \ , \ v_\pm=\frac{1}{\sqrt{2}}
\left[1-\frac{\xi_p \mp\mu}{\tilde E_0^\pm}\right]^{1/2}\ .
\end{eqnarray*}
The different solutions are orthogonal,
\[
\left<(\check{\hat \psi}_i)^\dagger \check{\hat
\psi}_k\right>\equiv {\rm Tr}[(\check{\hat \psi}_i)^\dagger
\check{\hat \psi}_k]=|C|^2 \delta_{ik}
\]
since $ \check A_2^\dagger  \check A_1 =\check A_4^\dagger \check
A_3 =0 $. The trace is taken over pseudo-spin and Nambu indexes.

If the coefficients $\check A^\pm$ are taken in the zero order
approximation in ${\bf k}$, the product $\alpha_n^+ \alpha_n^-$ in
Eq. (\ref{alpha+n-k}) contains the exponents $e^{-i\phi}$ and
$(k_x+ik_y)e^{-2i\phi}$ and vanishes after integration over the
momentum directions. Therefore, the basis functions $\zeta_n^\pm $
can be taken in zero approximation in ${\bf k}$ but the
coefficients $\check A^\pm$ need to be calculated up to the first
order terms in ${\bf k}$.


The corrections due to the condensate momentum can be written as
\begin{equation}
\check{\hat \psi}_\alpha =\check{\hat \psi}_\alpha^{(0)}+
\sum_{\beta \ne \alpha}B_{\alpha \beta}\check{\hat
\psi}_\beta^{(0)} \ . \label{corr-psi}
\end{equation}
Equations (\ref{matrix-eq}) - (\ref{H1}) give
\begin{eqnarray}
\delta \tilde E_{\alpha}=\frac{\left<(\check{\hat
\psi}_\alpha)^\dagger \check{\hat H}_1 \check{\hat
\psi}_\alpha\right>}{|C|^{2} }  , \;
B_{\alpha\beta}=\frac{\left<(\check{\hat \psi}_\beta)^\dagger
\check{\hat H}_1 \check{\hat \psi}_\alpha\right>}{|C|^{2} (\tilde
E_\alpha -\tilde E_\beta)}. \ \label{corr-B}
\end{eqnarray}
Corrections to energies are
\[
\delta \tilde E_{1,2}= \frac{{\bf p}{\bf k}}{2 p}\frac{d\xi_p}{dp}
\equiv E_D\ , \; \delta \tilde E_{3,4}= -\frac{{\bf p}{\bf k}}{2
p}\frac{d\xi_p}{dp}\equiv -E_D
\]
which is the usual normal-state Doppler shift. We have
$B_{12}=B_{21}=B_{34}=B_{43}=0 $, while
$B_{13}=B_{31}=-B_{24}=-B_{42}$ and $B_{23}=B_{32}=B_{14}=B_{41}$
where
\begin{eqnarray*}
B_{13}=-\frac{i([{\bf p}\times {\bf k}]{\bf
z})}{2p}\frac{d\xi_p}{dp}\frac{(u_+v_- + v_+u_-)}{(\tilde E^+_0
-\tilde E^-_0)}\ , \\
B_{23}=-\frac{i([{\bf p}\times {\bf k}]{\bf
z})}{2p}\frac{d\xi_p}{dp}\frac{(u_+u_- - v_+v_-)}{(\tilde E^+_0
+\tilde E^-_0)}\ .
\end{eqnarray*}


The currents Eq. (\ref{currentN}) at layer $n$ contains the
product of $\zeta^{+}_n \zeta^{-*}_n =(\xi_p/v_Fp)e^{i(N-1)\phi}$
which is independent of the layer number, i.e., of the distance
from the surface, and the products $[(\tilde E\pm \tilde
\mu)/t]\zeta^{+*}_n \zeta^+_n \propto (v_Fp/t)^{2(n-1)}$ and
$[(\tilde E\pm\tilde \mu)/t]\zeta^{-*}_n \zeta^-_n \propto
(v_Fp/t)^{2(N-n)}$ which decay as functions of the distance from
the surfaces. All these terms are of the order of $E/t$. We shall
see, however, that it is the constant term that gives the main
contribution to the total current through the sample, ${\bf I}=d
\sum_{n=1}^N {\bf j}(n)$. Using $ \sum_{\bf p}=\int (dp/d\xi_p)
p\, d\xi_p/2\pi\hbar $ we find for the current per unit sample
width
\begin{eqnarray}
{\bf I} &=&e d N  {\bf k}\int \frac{ \xi_p\, d\xi_p}{2\pi\hbar}
|C|^2\nonumber \\
&\times &\left[ \frac{1}{\xi_p} -
\left(\tanh\frac{E_0^+}{2T}+\tanh\frac{E_0^-}{2T}
\right)\frac{(u_+u_--v_+v_-)^2}{(\tilde E^+_0+\tilde E^-_0)}
\right.\nonumber \\
&-&\left. \left(\tanh\frac{E_0^+}{2T}-\tanh\frac{E_0^-}{2T}
\right)
\frac{(u_+v_- +v_+u_-)^2}{(\tilde E^+_0-\tilde E^-_0)}\right.\nonumber \\
&-&\left.\frac{1}{4T}\left(\cosh^{-2}\frac{E_0^+}{2T} +
\cosh^{-2}\frac{E_0^-}{2T}\right)\right]\ . \label{Current-full1}
\end{eqnarray}

To obtain this expression we had to regularize Eq.
(\ref{currentN}) which diverges for large $\xi_p$. The
regularization is described in detail in Ref.
\cite{KopninSonin10}. In brief, we subtract the normal current
which is obtained from the current operator taken at energies much
higher than $\Delta$ and $T$. For $\xi_p \gg \Delta, T$ one has
$E^+_0=\xi_p-\mu$, $E^-_0=\xi_p+\mu$, $u =1$, and $v=0$.
Therefore, the diverging part of Eq. (\ref{currentN}) is
\begin{eqnarray}
{\bf I}^{(\infty)}=-d e N {\bf k}\int \frac{\xi_p\,
d\xi_p}{2\pi\hbar} |C|^2 \frac{1}{\xi_p} \ . \label{current-norm}
\end{eqnarray}
This contributes to the normal current which, of course, turns to
zero in the end. Indeed, for $\Delta=0$ when the particle and hole
channels separate, the corrections to $\check A^\pm$ simply
correspond to the full shift of the momentum ${\bf p}\to {\bf
p}\pm {\bf k}/2$ in the particle (hole) wave functions. As a
result, the normal current vanishes after the momentum integration
over the entire Brillouin zone \cite{KopninSonin10}. After
subtracting the zero normal current, we arrive at Eq.
(\ref{Current-full1}).

For low temperature $T\ll |\Delta|$, the last two lines in Eq.
(\ref{Current-full1}) turn to zero. The total current thus becomes
\begin{eqnarray}
{\bf I} =d e N  {\bf k}\int \frac{\xi_p\, d\xi_p}{2\pi\hbar}|C|^2
\left[ \frac{1}{\xi_p} - \frac{2(u_+u_- -v_+v_-)^2}{(\tilde
E^+_0+\tilde E^-_0)}\right] \label{current12-0}
\end{eqnarray}
which is similar to the result obtained in Ref.
\cite{KopninSonin10}. For $\mu=0$ we have
\begin{eqnarray*}
{\bf I} =de N {\bf k}\int_0^\infty \frac{
d\xi_p}{2\pi\hbar}|C|^2\left(1
 - \frac{\xi_p^3}{\tilde E_0^3}\right)
\end{eqnarray*}

For large $N$ one can consider $p$ as a slow function as compared
to $\xi_p$. This is equivalent to the assumption that $ d\left[
\xi_p \left(1- v_F^2p^2/t^2\right)\right]/ dp =\left(1-
v_F^2p^2/t^2\right)(d\xi_p/dp)$ i.e., that $ \left(1-
v_F^2p^2/t^2\right) \gg (2/N) (v_F^2p^2/t^2) $. Since $\xi_p \sim
\Delta$ we have
\[
1- v_F^2p^2/t^2 = 1-
\left(\Delta/t\right)^\frac{2}{N}=(2/N)\ln(t/\Delta)
\]
which holds for $N\gg \ln(t/\Delta) $. Therefore, the above
condition is satisfied within the logarithmic approximation. Note
that neglecting the terms $\zeta^{+*}_n \zeta^+_n $ and
$\zeta^{-*}_n \zeta^-_n $ in Eq. (\ref{currentN}) that decay away
from the surfaces is also legitimate within the same logarithmic
approximation $\ln(t/\Delta)\gg 1$. Integrating by parts and using
that the integral is determined by $\xi_p \sim \Delta$ we find
\begin{eqnarray*}
{\bf I} =\frac{e N\Delta^2 {\bf k}}{\pi\hbar}\int_0^\infty
\left(1-\frac{v_F^2p^2}{t^2}\right)\frac{\xi_p }{\tilde E_0^3}\,
d\xi_p =\frac{2 e \Delta \ln (t/\Delta) k_x }{\pi\hbar } .
\end{eqnarray*}
The total current does not depend on the sample thickness $Nd$ as
it should be for the surface superconductivity. The critical
current is determined by ${\rm max}(k_x) \sim \xi_0^{-1}$ where
the coherence length is\cite{KopninHeikkilaVolovik11} $\xi _0 \sim
\hbar /p_{\rm FB}=\hbar v_{F}/t$,
\[
I_c\sim e \Delta \ln (t/\Delta) p_{\rm FB}\ .
\]

For nonzero $\mu$ we find in the same way as in
Ref.\cite{KopninSonin10}
\begin{eqnarray}
{\bf I}&=&\frac{ e \ln (t/\Delta) {\bf  k} }{\pi\hbar }\left[
\sqrt{|\mu|^2 +|\Delta|^2} \right. \nonumber \\
&&+\left. \frac{|\Delta|^2}{|\mu|}\ln\left(\frac{|\mu|+
\sqrt{|\mu|^2 +|\Delta|^2}}{|\Delta|}\right)\right]\ .
\label{curr-mu0}
\end{eqnarray}
Recall that Eq. (\ref{curr-mu0}) holds for $T\ll |\Delta|$. As
distinct from the case of intrinsic superconductivity in graphene
considered in Refs.
\cite{KopninSonin10,CastroNeto05,KopninSonin08}, the surface
superconductivity gap $|\Delta|$ is suppressed by doping
\cite{KopninHeikkilaVolovik11}, such that both $|\Delta|$ and
$T_c$ vanish as $\mu$ reaches the critical level $\mu_c= 2T_{c0}$.

To conclude, we have calculated the zero-temperature supercurrent
for the surface superconductivity of a flat-band multilayered
rhombohedral graphene. The supercurrent is finite despite the
absence of dispersion of the excitation spectrum. The critical
current is proportional to the zero-temperature gap, i.e., to the
superconducting critical temperature and to the size of the flat
band in the momentum space. Nonzero surface supercurrent can be
responsible for the small Meissner effect and for the sharp drop
in resistance seen in experiments on graphite
\cite{Kopelevich01,Esquinazi08}. The enhanced superconducting
density has been reported on twin boundaries in
Ba(Fe$_{1-x}$Co$_x$)$_2$As$_2$ \cite{Moler2010}. This observation
can also be considered as indications towards surface
superconductivity described by our theory.


\acknowledgements

I thank G. Volovik for helpful discussions. This work is supported
by the Academy of Finland Centers of excellence program
2006--2011, by the Russian Foundation for Basic Research (grant
09-02-00573-a), and by the Program ``Quantum Physics of Condensed
Matter'' of the Russian Academy of Sciences.

\end{document}